\documentclass[twocolumn, prl, aps, superscriptaddress,longbibliography,showpacs,amsmath,amssymb,floatfix]{revtex4-1}   
\usepackage{verbatim}
\usepackage{amsfonts}
\usepackage{bbm}
\usepackage{subfigure}
\usepackage{tipa}
\usepackage{color}
\usepackage{epstopdf}
\usepackage{epsfig}
\usepackage{subfigure}
\usepackage{mathrsfs} 
\usepackage{amsmath,amssymb,amsfonts}
\usepackage{graphicx}
\usepackage{dcolumn}
\usepackage{bm}
\usepackage{braket}
\usepackage{amssymb}
\usepackage[colorlinks,linkcolor=blue,anchorcolor=blue,citecolor=blue,urlcolor=blue]{hyperref}
\makeatletter

\newcommand{\Rmnum}[1]{\expandafter\@slowromancap\romannumeral #1@}

\makeatother

\begin{document}
\title{Unilateral Criticality and Phase Transition in the Cavity-Ising Model}
\author{Zeyu Rao}
\affiliation{Key Laboratory of Quantum Information, University of Science and Technology of China, Hefei 230026, China}
\affiliation{Anhui Province Key Laboratory of Quantum Network,
University of Science and Technology of China, Hefei 230026, China}
\author{Xiaoshui Lin}
\affiliation{Key Laboratory of Quantum Information, University of Science and Technology of China, Hefei 230026, China}
\affiliation{Anhui Province Key Laboratory of Quantum Network,
University of Science and Technology of China, Hefei 230026, China}
\author{Xiwang Luo}
\email{xwluo@ustc.edu.cn}
\affiliation{Key Laboratory of Quantum Information, University of Science and Technology of China, Hefei 230026, China}
\affiliation{Anhui Province Key Laboratory of Quantum Network,
University of Science and Technology of China, Hefei 230026, China}
\affiliation{Hefei National Laboratory, University of Science and Technology of China, Hefei 230088, China}
\affiliation{Synergetic Innovation Center of Quantum Information and Quantum Physics, University of Science and Technology of China, Hefei 230026, China}
\author{Guangcan Guo}
\affiliation{Key Laboratory of Quantum Information, University of Science and Technology of China, Hefei 230026, China}
\affiliation{Anhui Province Key Laboratory of Quantum Network,
University of Science and Technology of China, Hefei 230026, China}
\affiliation{Hefei National Laboratory, University of Science and Technology of China, Hefei 230088, China}
\affiliation{Synergetic Innovation Center of Quantum Information and Quantum Physics, University of Science and Technology of China, Hefei 230026, China}
\author{Han Pu}
\email{hpu@rice.edu}
\affiliation{Department of Physics and Astronomy, and Smalley-Curl Institute, Rice University, Houston, Texas 77251-1892, USA}
\author{Ming Gong}
\email{gongm@ustc.edu.cn}
\affiliation{Key Laboratory of Quantum Information, University of Science and Technology of China, Hefei 230026, China}
\affiliation{Anhui Province Key Laboratory of Quantum Network,
University of Science and Technology of China, Hefei 230026, China}
\affiliation{Hefei National Laboratory, University of Science and Technology of China, Hefei 230088, China}
\affiliation{Synergetic Innovation Center of Quantum Information and Quantum Physics, University of Science and Technology of China, Hefei 230026, China}


\begin{abstract}
Superradiant phase transitions from cavity light-matter coupling have been widely explored across platforms. Here, we report a unilateral critical endpoint (UCEP) and a tricritical point (TCP) in the phase diagram of the cavity-coupled transverse Ising model with $\mathbb{Z}_2$ symmetry.
At zero temperature, we demonstrate that this model hosts three phases separated by two second-order and one first-order transitions. These lines intersect at a TCP and a UCEP, the latter not captured by existing phase-transition paradigms. The UCEP displays one-sided criticality: approaching the point from one side, the system behaves as a second-order transition, while from the other side it is first-order. Correspondingly, two order parameters, respectively, undergo the first- and the second-order phase transitions at the same point. We construct a minimal description of UCEP with the density of the free energy $f = c_{1}(\tilde{\alpha}^{2}+c_{2})+(\tilde{\alpha}^{2}+c_{2})^{2}\ln{\vert\tilde{\alpha}^{2}+c_{2}\vert}$, with the UCEP at $(c_{1},c_{2})=(1/e,0)$ and $\tilde{\alpha}$ being the order parameter. We further map the finite-temperature phase diagram and perform a symmetry analysis. By unifying first- and second-order signatures in a single, direction-dependent endpoint, the UCEP introduces a qualitatively new class of phase transition and may have applications in fields such as quantum measurement and quantum sensing. This work also provides an intriguing platform for exploring novel critical phenomena in cavity-coupled many-body systems with or without dissipation.

\end{abstract}
\maketitle

The Dicke model \cite{Dicke1954Coherence,WangY1974PhaseTransition, Hioe1973PhaseTransitions,Tavis1968Exact,Duncan1974Effect,popov1982behavior,popov1988functional,Vukics2012Adequacy, rzazewski1975phase}, which is an ensemble of two-level systems coupled to a quantized cavity field, has been extensively studied \cite{baden2014realization, Nagy2010Dicke, 
dimer2007proposed, 
baumann_dicke_2010,Emary2003Chaos,Klaus1973On,Emary2003Quantum,Bakemeier2012Quantum,GROSS1982Superradiance,Nagy2011Critical,Nagy2016Critical,Baumann2011Exploring} and experimentally realized across multiple platforms, including ultracold atoms \cite{baumann_dicke_2010, klinder2015dynamical,Zhang2017Quantum, zhang2018dicke,dimer2007proposed,ritsch2013cold}, trapped ions \cite{Safavi2018Verification}, superconducting circuits \cite{lamata2017digital,mezzacapo2014digital,langford2017experimentally}, nuclear magnetic resonance \cite{Chen2021Experimental}, and solid-state systems \cite{Li2018ObservationOD,kono2024,doi:10.1126/sciadv.adt1691}.
While the conventional Dicke transition is second order, various extensions can alter its order \cite{fallas2023tricritical,
XuYoujiang2019Emergent,XuYouJiang2021Multicriticality}. Furthermore, a superradiant phase transition can also be realized with a single atom interacting with a cavity field, under the limit that the ratio of the frequency of the optical field and that of the atomic transition approaches zero \cite{Hwang2015Quantum, Filippis2023Signatures, Grimaudo2023Quantum,Liu2017Universal,Puebla2016Excited,Lv2018Quantum,Xie2017quantum,Braak2011Integrability,ashhab2013superradiance}. It has been proven that the phase transitions in these models are well described by the Landau theory of phase transition \cite{Zhuang2021Universality}. As intriguing applications, the quantized cavity may be used to explore the enhanced superconductivity \cite{Curtis2019CavitySC, Schlawin2019CavitymediatedSC} and the breakdown of the quantized quantum Hall effect \cite{Felice2022Breakdown}.  

We report unilateral (one-sided) criticality in a cavity-coupled transverse Ising model. Approaching the critical point from one side, the system behaves as a second-order phase transition: one order parameter vanishes continuously, and the susceptibility and entanglement entropy diverge logarithmically. From the other side, it is first-order, with another order parameter exhibiting a finite jump, and the divergence disappears. Our key findings are as follows. (I) At zero temperature, this model with $\mathbb{Z}_2$ symmetry exhibits three phases: the paramagnetic-superradiant (PSR) phase, paramagnetic-normal (PN) phase, and antiferromagnetic-normal (AFN) phase, separated by two second-order and one first-order transition lines. (II) These phase boundaries yield a tricritical point (TCP) and a unilateral critical endpoint (UCEP) that exhibits the above one-sided criticality,
which is a feature not captured by existing paradigms of phase transitions. Correspondingly, two order parameters respectively undergo first- and second-order phase transitions simultaneously at the same point. The minimal model capturing the UCEP (with one-sided logarithmic divergence) is given by the free energy density: $f = c_{1}(\tilde{\alpha}^{2}+c_{2})+(\tilde{\alpha}^{2}+c_{2})^{2}\ln{\vert\tilde{\alpha}^{2}+c_{2}\vert}$ (with UCEP at $c_{1}=1/e, c_{2}=0$, $e$ is Euler's number). (III) At finite temperature, the UCEP disappears and only the normal-superradiant transition remains. These results constitute a qualitatively new class  of phase transition and provide a platform for further exploration of cavity-coupled physics. We address this problem by exactly solving this model in the thermodynamic limit. 

\begin{figure}[htbp]
	\centering
 \includegraphics[width=0.48\textwidth]{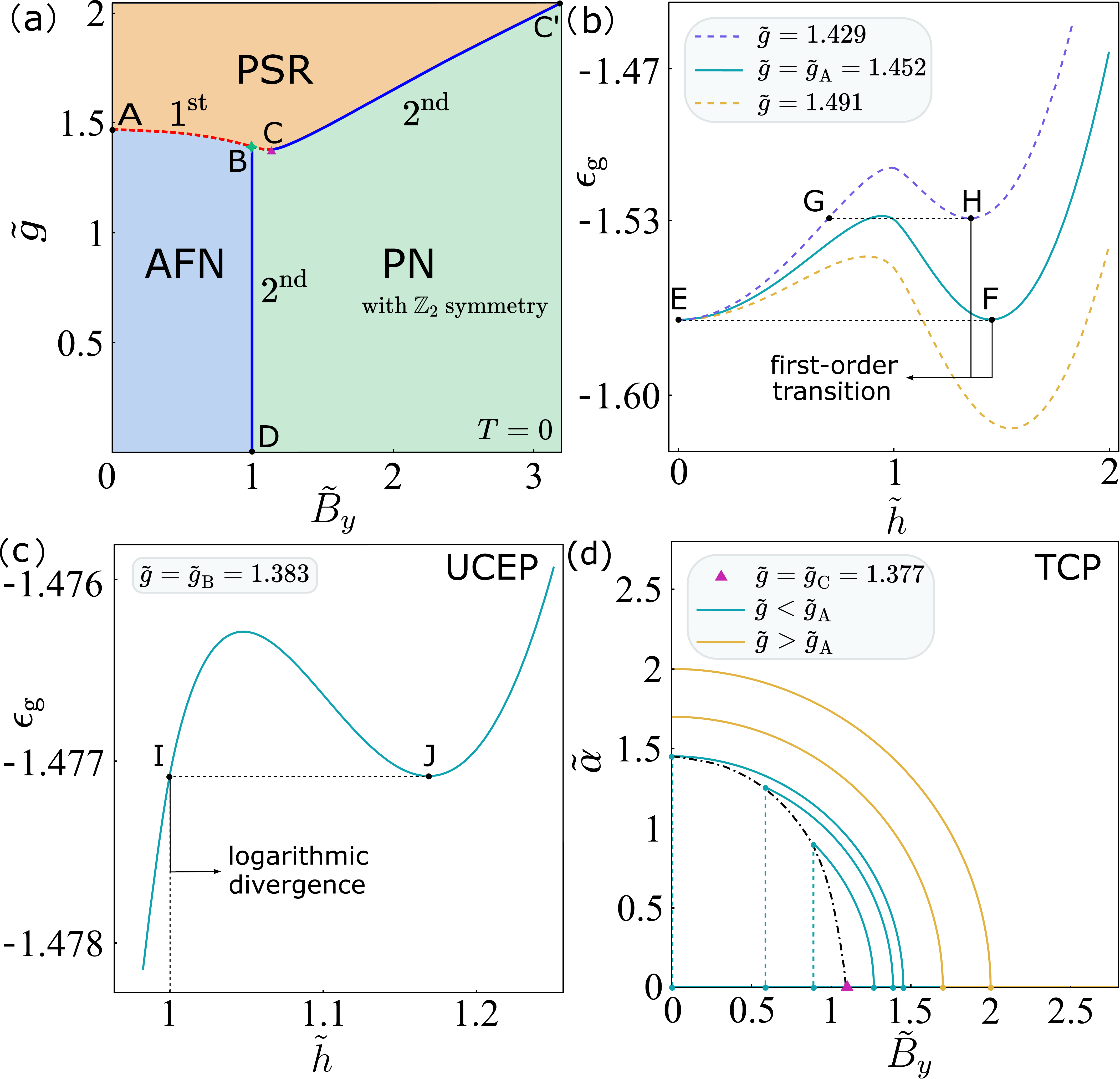}
\caption{(a) Phase diagram in $(\tilde{B}_y,\tilde{g})$ at $T=0$, with A ($0, 1.45249$), B ($1, 1.38288$) (green star), C ($1.10022, 1.37698$) (purple-pink triangle), and D ($1, 0.0$). Red dashed lines (AB, BC) and blue solid lines (BD, CC') are first- (1$^{\text{st}}$) and second-order (2$^{\text{nd}}$) transitions, respectively. (b) Ground-state energy $\epsilon_{\text{g}}(\tilde{h})$ for  $\tilde{g}=1.429,1.452,1.491$, representing an example of the first-order transition (line AB), point A, and the PSR phase, respectively. (c) Unilateral critical endpoint: points I (nonanalytic) and J (a local minimum) have the same value of $\epsilon_{\text{g}}$. As approaching point I,  $\epsilon_{\text{g}}''(\tilde{h})$ logarithmically diverges. (d) Superradiant strength $\tilde{\alpha}$ versus $\tilde{B}_{y}$. For $\tilde{g}>\tilde{g}_{\text{A}}$, only second-order phase transitions occur, while for $\tilde{g}<\tilde{g}_{\text{A}}$, it undergoes a first-order transition followed by a second-order transition. As $\tilde{g}$ decreases, the quarter circular arc shrinks to the purple-pink triangle point (tricritical point).}
\label{fig-fig1}
\end{figure}
{\it Moodel and Hamiltonian}: We consider the transverse one-dimensional Ising model in a cavity, which can be regarded as the simplest many-body system coupled to a cavity. Previous studies of this model with cavity loss have shown the existence of a first-order transition, but the criticality is not clarified ~\cite{luo2016dynamic,zhang2014quantum,Gammelmark2011Phase,Roche2025bound,Roche2025Linear,langheld2024quantum}. The Hamiltonian for this system reads
\begin{equation}
H = \omega\hat{a}^{\dagger}\hat{a} + \sum_i  J \sigma_i^z \sigma_{i+1}^z + \frac{g}{\sqrt{N}}(\hat{a}+\hat{a}^{\dagger}) \sigma_i^x + B_y\sigma_i^y,
\label{eq-cavity_ising}
\end{equation}
where $a$ is the annihilation operator of the cavity field, $\sigma_i^q$ ($q = z$, $x$, $y$) the atomic Pauli operators, $\omega$ the cavity photon frequency, $J$ the Ising interaction strength between neighboring atoms (we focus on antiferromagnetic coupling with $J>0$ \cite{SignofJ}), $g$ the coupling strength between the cavity and the atoms, and $B_y$ the transverse Zeeman field along the $y$-direction. Various ways to realize this model have been discussed in the literature \cite{Whitlock2017Simulating,Nguyen2018Towards,Signoles2021Glassy,Duan2003Controlling,jepsen2020spin,Zhao2019Engineered,Pai2005Superfluid,Davoudi2020Towards}, and the generalization of this model to network models is discussed in Ref.~\cite{Bazhenov2021mean}. 

{\it Partition function}: The partition function at temperature $T$ can be obtained using the method developed in Refs.~\cite{WangY1974PhaseTransition, Hioe1973PhaseTransitions}, see also Refs.~\cite{Klaus1973On, Dicke1954Coherence, Zhuang2021Universality,Chen2008Numerically}:
\begin{align}
    Z& =\sum_{s}\int \frac{\text{d}^{2}\alpha}{\pi} \,\langle s,\alpha\vert e^{-\beta H}\vert s,\alpha\rangle = \int \frac{\text{d}^{2}\alpha}{\pi}\,\mathrm{Tr}_{s}(e^{-\beta  \mathcal{H}(\alpha)}) , \label{zz}\\ 
\mathcal{H}& =\omega \vert \alpha\vert^{2}+\sum_i  \left[J \sigma_i^z \sigma_{i+1}^z + h \sigma_i^x + B_y\sigma_i^y \right], \label{hh}
\end{align}
where $\beta \equiv 1/(k_B T)$, $h \equiv \frac{2g\operatorname{Re}(\alpha)}{\sqrt{N}}$, $|\alpha\rangle$ represents photon coherent state, $|s\rangle =\Pi_i \otimes |s_i \rangle $ atomic spin product state, and the trace is carried over the whole spin space. The integral in Eq.~\eqref{zz} can be carried out using the saddle point approximation, which becomes exact in the thermodynamic limit \cite{Zhuang2021Universality}. The free energy follows as $F = -\beta^{-1} \ln Z$. At zero temperature, $F$ reduces to the ground-state energy $E_g$. 

{\it Symmetry}: 
The Hamiltonian $H$ is $\mathbb{Z}_2$ symmetric under $\hat{a}\rightarrow -\hat{a}$, $ \sigma_i^z \rightarrow -\sigma_i^z$, $\sigma_i^y \rightarrow \sigma_i^y$, $\sigma_i^x \rightarrow -\sigma_i^x$, generated by $\mathcal{A} = e^{i\pi \hat{a}^\dagger \hat{a}} \prod_{i}\otimes \sigma_i^y$ with
$\mathcal{A} H \mathcal{A}^{-1} = H$.
Although $H$ features a set of four independent parameters $\{ \omega, J, g, B_y \}$, the ground-state phases are determined by two dimensionless quantities:
$
    \tilde{g} \equiv 2\sqrt{2}g/\sqrt{\pi J \omega}\,, \quad \tilde{B}_y \equiv B_y/J . 
$
Fig.~\ref{fig-fig1}(a) shows the zero-temperature phase diagram in the $(\tilde{B}_{y},\tilde{g})$ plane with three phases: PSR, PN, and AFN.
The PN phase is $\mathbb{Z}_2$ symmetric, while both PSR and AFN break the $\mathbb{Z}_2$ symmetry. Two order parameters characterize these phases
$
    \alpha \equiv \langle \hat{a} \rangle\,, \quad m_s \equiv \sum_i \langle (-1)^i \sigma_i^z \rangle /N ,
$
which arise from the Dicke normal-superradiant and the Ising paramagnetic-antiferromagnetic transitions, respectively. We examine these three phases in several limits \cite{Sachdev2011Book}. Firstly, in the AFN phase with $\tilde{g} \sim 0$ and $\tilde{B}_y = 0$, the two degenerate ground states should be 
$
|\cdots \uparrow\downarrow \uparrow\downarrow\uparrow\downarrow \cdots \rangle \otimes |0\rangle$, $ |\cdots \downarrow \uparrow\downarrow\uparrow\downarrow\uparrow \cdots \rangle \otimes |0\rangle.
$
Thus, the order parameters are given by $|m_s| \neq 0$ and $\alpha = \langle \hat{a} \rangle = 0$, with long-range spin correlation $\lim_{l\rightarrow \infty} \langle \sigma_i^z \sigma_{i+l}^z\rangle \rightarrow \text{finite}$. Secondly, in the PN phase when $\tilde{g}\sim 0$ and $\tilde{B}_y \gg J$, the ground state is non-degenerate
$
|\cdots \rightarrow \rightarrow\rightarrow\rightarrow\rightarrow \cdots \rangle \otimes |0\rangle,
$
yields $\alpha = \langle \hat{a} \rangle = 0$ and $m_s  =  0$, and short-range spin correlation $\langle \sigma_i^z \sigma_{i+l}^z\rangle \sim \exp(-l/\xi)$, where $\xi$ is the correlation length. Finally, in the PSR phase, the two ground states should be 
$
|\cdots \nearrow \nearrow \nearrow \cdots \rangle \otimes |\alpha\rangle, \quad 
|\cdots \swarrow \swarrow \swarrow \cdots \rangle \otimes |-\alpha\rangle,
$
where $|\nearrow \rangle$ and $|\swarrow \rangle$ denote atomic spin state polarized in the transverse plane. The corresponding order parameters are $m_s = 0$ and $\alpha \propto \sqrt{N(g-g_c)}$, where $g_c$ is the critical coupling strength, with short-range spin correlation $\langle \sigma_i^z \sigma_{i+l}^z\rangle \sim \exp(-l/\xi)$. Since both AFN and PSR phases break the $\mathbb{Z}_2$ symmetry, the transition between them is necessarily of first order. Otherwise, two sequential spontaneous symmetry breakings should occur along the PN $\rightarrow$ AFN $\rightarrow$ PSR, which is not allowed since the system possesses the sole $\mathbb{Z}_2$ symmetry. Now we discuss the phase diagram, particularly the four special points, labeled A, B, C, and D in Fig.~\ref{fig-fig1}(a). 

{\it Ground-state phase diagram}: (1) Consider $B_y = 0$.  Jordan-Wigner transformation 
will transform $\mathcal{H}$ in Eq.~\eqref{hh} into a spinless fermion model \cite{Barouch1970Statistical,parkinson2010introduction,franchini2017introduction},
which can be solved analytically by writing $\mathcal{H}$ in the momentum space \cite{Sachdev2011Book,Barouch1970Statistical,parkinson2010introduction,franchini2017introduction}.
In this way, the cavity-Ising model is mapped to a model in which $p$-wave superconducting fermions couple to a cavity, with ground-state energy
\begin{equation}
    E_\text{g}=\omega\alpha^{2}-\frac{N}{2\pi}\int_{-\pi}^{\pi}dk \, \Omega_k,
    \label{eq-Eg1}
\end{equation}
with the spectra $\Omega_k = \sqrt{(h-J\cos{k})^{2}+J^2\sin^{2}{k}}$. We solve Eq.~\eqref{eq-Eg1} exactly and introduce a normalized dimensionless energy density as $\epsilon_\text{g} = \pi E_\text{g}/(2JN)$
\begin{equation}
    \epsilon_\text{g}  = \frac{\tilde{h}^2}{\tilde{g}^{2}} - |\tilde{h}-1\vert \text{E}(-\frac{4\tilde{h}}{(\tilde{h}-1)^{2}}) \,,
\label{equation-eg1}
\end{equation}
where $\tilde{h} \equiv h/J$, and E$(x)$ is the Elliptic function of second kind. In Fig.~\ref{fig-fig1}(b), we plot $\epsilon_\text{g}$ as a function of $\tilde{h}$ for several different values of $\tilde{g}$. $\epsilon_{\text{g}}(\tilde{h})$ is an even function of $\tilde{h}$, hence we only plot on the positive $\tilde{h}$-axis. These curves depict typical first-order normal-superradiant phase transitions. Point A in Fig.~\ref{fig-fig1}(a), can be obtained by 
\begin{equation}
   \left. \epsilon_{\text{g}}(\tilde{h}) \right|_{\tilde{h}=\tilde{h}_0} \!  = \left. \epsilon_{\text{g}}(\tilde{h}) \right|_{\tilde{h}=0}, \,\,\, \left. \epsilon_{\text{g}}'(\tilde{h}) \right|_{\tilde{h}=\tilde{h}_0} \! =0, \,\,\, \tilde{h}_0 > 0, 
    \label{equation-eg2}
\end{equation}
 with the critical value $\tilde{g}_{A}=1.45957$. 

(2) When $g = 0$, the cavity decouples and the transverse Ising model undergoes the second-order phase transition at $B_y = J$ (i.e., $\tilde{B}_y=1$) from the antiferromagnetic phase ($B_y < J$) to the paramagnetic phase ($B_y > J$), marking point D in Fig.~\ref{fig-fig1}(a). In the normal phase ($\alpha = 0$), the coupling $g$ is not important, producing the vertical second-order critical line BD. 

(3) For $B_y \ne 0$ and $g \ne 0$, we perform a spin rotation about the $z$-axis under the unitary operator $\mathcal{R} = \exp(\frac{i\sigma_i^{z}\phi}{2})$ with $\tan(\phi)=\frac{B_{y}\sqrt{N}}{2\alpha g}$, leading to  
$
    \mathcal{H} = \omega\alpha^{2}+\sum_{i} \left[J\sigma_{i}^{z}\sigma_{i+1}^{z}+\sqrt{\frac{4\alpha^{2}g^{2}}{N}+B_{y}^{2}} \, \sigma_{i}^{x} \right]\,,
$
which has the same form as Eq.~\eqref{hh}. Hence, the renormalized energy density in the same form as Eq.~\eqref{equation-eg1} with
a new $\tilde{h} \equiv \sqrt{\tilde{B_{y}}^{2}+\tilde{\alpha}^{2}}$
and $ \tilde{\alpha}^{2}\equiv \frac{4\alpha^{2}g^{2}}{NJ^{2}}$ can be obtained. Consequently, the phase diagram is fully determined by the two dimensionless parameters: $\tilde{g}$ and $\tilde{B}_{y}$.

(4) When $g, B_y \gg J$, the Ising coupling between neighboring spins is negligible, and the Hamiltonian is reduced to the standard Dicke model with $\mathcal{H} =\omega\alpha^{2}+ h S_z + B_y S_y$, $h=\frac{2g\alpha}{\sqrt{N}}$, where $S_q = \sum_i \sigma_i^q$, $q = x, y, z$ and $\alpha \in \mathbb{R}$. The second-order superradiant transition occurs at $g_c = \sqrt{\omega B_y}$ \cite{WangY1974PhaseTransition, Hioe1973PhaseTransitions, Zhuang2021Universality}. As a result, along the phase boundary BC', there should exist a TCP at which the normal-superradiant transition changes from first- to second-order. 
Multicriticality has been explored in different cases \cite{Chang1973Generalized,Hankey1972Geometric,Griffiths1975Phase,Tuthill1975Renormalization,Riedel1972Scaling,CHANG1992Differential,Griffiths1970Thermodynamics,Soriente2018Dissipation,Casta2012Universal,Overbeck2017Multicritical}. From Fig.~\ref{fig-fig1}(d), the circular arc shrinks to a TCP as $\tilde{g}$ decreases. At the TCP,  both the first- and second-order derivatives of $\epsilon_\text{g}$ vanish, leading to: 
\begin{equation}
\epsilon_{\text{g}}'(\tilde{h}) = 0, \quad 
\epsilon_{\text{g}}''(\tilde{h}) = 0, \quad \tilde{h} > 0,
\end{equation}
from which we have $\tilde{g}_{C}=1.37698$ and $\tilde{B}_{y}^\text{C}=1.10022$. With $\tilde{\alpha}^2 \ll 1$, we have the expansion
$\epsilon_{\text{g}}=-1.471+0.086\tilde{\alpha}^{6}-0.156\tilde{\alpha}^{8}$.
The coefficients of both the $\tilde{\alpha}^{2}$ and $\tilde{\alpha}^{4}$ terms vanish, matching the Landau theory for the tricritical point \cite{landau2013statistical}. 
\begin{figure}[htpb]
\centering
\includegraphics[width=0.485\textwidth]{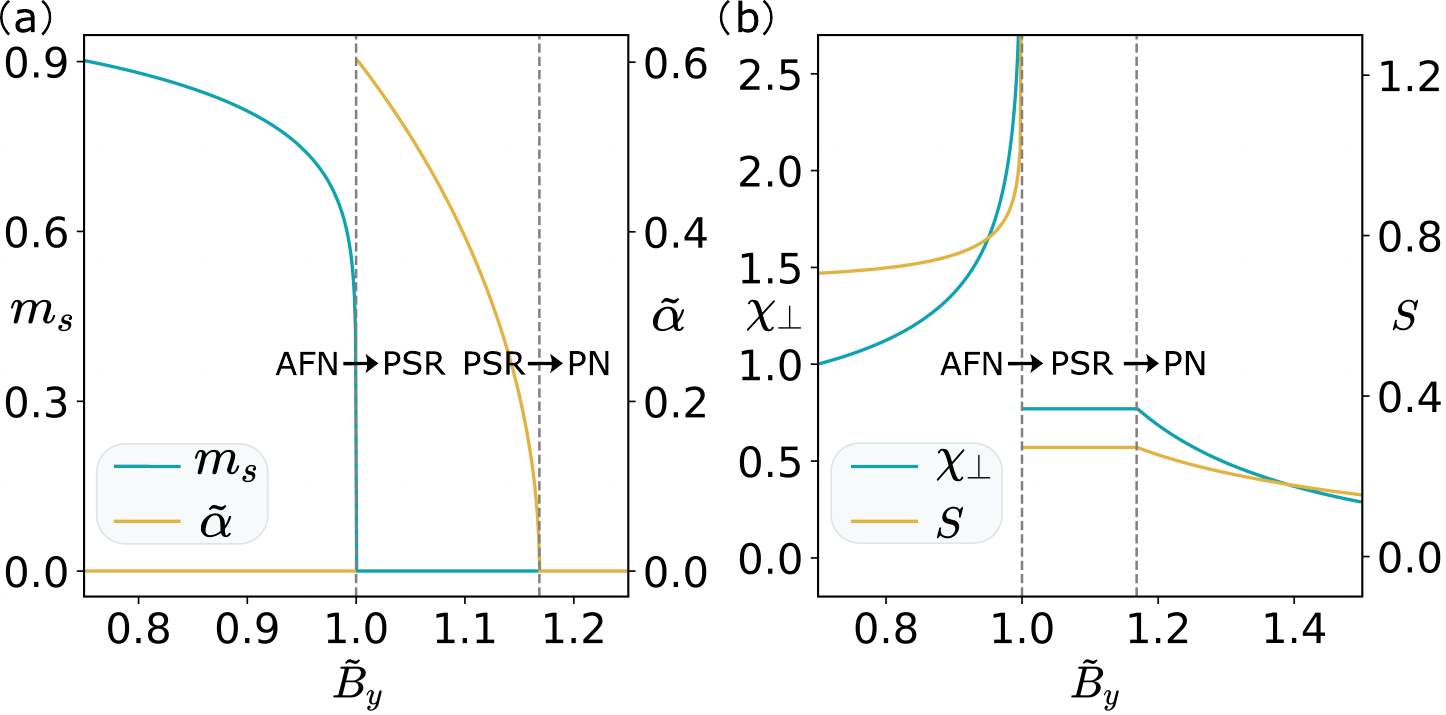}
\caption{(a) The order parameters across the UCEP, with $\tilde{g}=\tilde{g}_{\text{B}}=1.38288$, with second- ($m_{s}$) and first-order ($\tilde{\alpha}$) phase transitions simultaneously happening at $\tilde{B}_{y}=1$. (b) The transverse susceptibility and the entanglement entropy exhibit logarithmic divergence when $\tilde{B}_{y}\rightarrow 1^{-}$,  while remaining finite when $\tilde{B}_{y}\rightarrow1^{+}$, with another second-order transition at $\tilde{B}_{y}=1.16845$.}
\label{fig-fig2}
\end{figure}
\begin{figure}[htbp]
	\centering
 \includegraphics[width=0.46\textwidth]{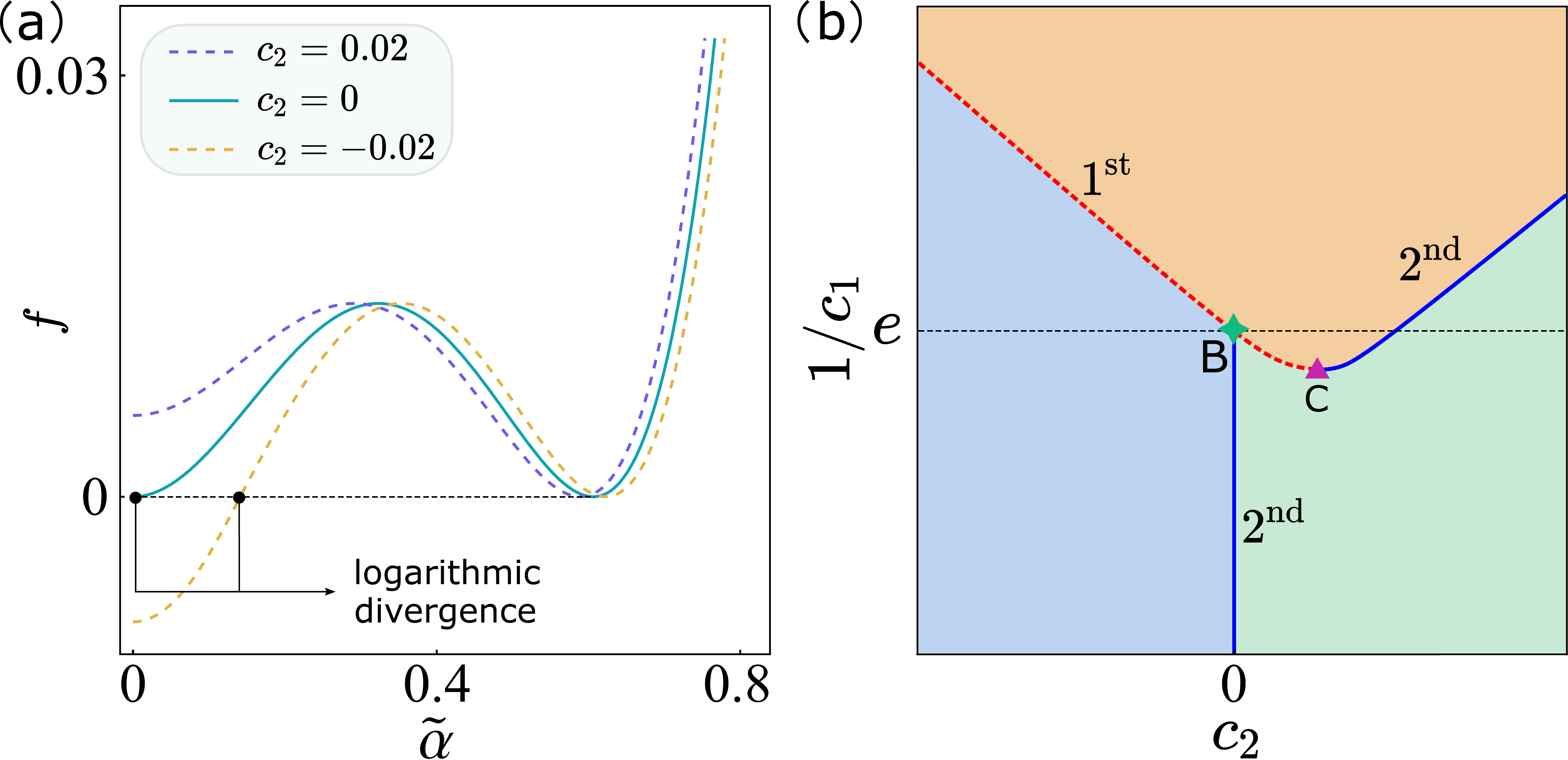}
\label{fig-fig3}
\caption{(a) The density of free energy in Eq.~\eqref{eq-UCEP} with $c_{1}=1/e$. The black dots denote the nonanalytic points. (b) The schematic phase diagram of Eq.~\eqref{eq-UCEP}. The red dashed and blue solid lines represent first- and continuous phase transitions, respectively; points B and C represent the UCEP and the TCP. }
\label{fig-fig3}
\end{figure}

(5) The UCEP (point B) arises from 
\begin{equation}
   \left. \epsilon_{\text{g}}(\tilde{h}) \right|_{\tilde{h}=\tilde{h}_0} \!  = \left. \epsilon_{\text{g}}(\tilde{h}) \right|_{\tilde{h}=1}, \,\,\, \left. \epsilon_{\text{g}}'(\tilde{h}) \right|_{\tilde{h}=\tilde{h}_0} \! =0, \,\,\, \tilde{h}_0 > 0, 
\end{equation}
which yields points I and J in Fig.~\ref{fig-fig1}(c), and $\tilde{g}_{B}=1.38288$. At $\tilde{h}=1, \tilde{B}_{y}=1$, when $\tilde{\alpha}^{2}\ll 1$, the expansion gives 
\begin{equation}
    \epsilon_{\text{g}}=\frac{1-2\tilde{g}^{2}}{\tilde{g}^{2}}+\frac{2-\tilde{g}^{2}}{2\tilde{g}^{2}}\tilde{\alpha}^{2}+\frac{1}{16}\ln{(\tilde{\alpha}^{2})}\tilde{\alpha}^{4}.
    \label{eq-point_B}
\end{equation}
Although the critical endpoints \cite{Fisher1990Universality,Szydlowska2023Ferroelectric,poole1992phase} can be explained by Landau's theory \cite{Szydlowska2023Ferroelectric}, and have previously been studied in different materials, such as $\text{UTe}_{2}$ \cite{wu2025Quantum}, $\text{Sr}_{3}\text{Ru}_{2}\text{O}_{7}$ \cite{grigera2001magnetic,rost2009entropy}, and supercooled silicon \cite{vasisht2011liquid}, the UCEP identified here does not fit existing paradigms of phase transition. Unlike critical endpoints previously studied, where the jump of the first-order phase transition tends to zero as one approaches the critical endpoint, while in our case, the jump of $\tilde{\alpha}$ across point B remains finite. In Fig.~\ref{fig-fig2}(a), when crossing the UCEP, the two order parameters $m_{s}$ and $\tilde{\alpha}$ simultaneously undergo second- and first-order phase transitions, respectively, and from Fig.~\ref{fig-fig2}(b), both the transverse susceptibility and the entanglement entropy of the spin systems exhibit logarithmic divergence ($\chi_{\perp}\sim -\frac{1}{2}\ln{\vert \tilde{B}_{y}-1\vert}$ and $S\sim -\frac{1}{12}\ln{\vert \tilde{B}_{y}-1\vert}$) only when $\tilde{B}_{y}\rightarrow 1^{-}$, but remain a finite value without divergence when $\tilde{B}_{y}\rightarrow 1^{+}$. Both the dual-transition property and the one-sided divergence can be summarized into one feature named unilateral criticality: the critical behavior on one side acts as a second-order phase transition. In contrast, on the other side, it acts as a first-order transition. The specific calculation of $\chi_{\perp}$ and $S$ and technical details are in SM \cite{suppRao}.

The minimal description of the UCEP is given by the density of the free energy:
\begin{equation}
    f = c_{1}(\tilde{\alpha}^{2}+c_{2})+(\tilde{\alpha}^{2}+c_{2})^{2}\ln{\vert\tilde{\alpha}^{2}+c_{2}\vert},
    \label{eq-UCEP}
\end{equation}
where $c_{1}$, $c_{2}$ are parameters and $\tilde{\alpha}$ is the order parameter. To match our cavity-Ising model, the expansion of $\tilde{h}=\sqrt{(\tilde{B}_{y}+\delta\tilde{B}_{y})^{2}+\tilde{\alpha}^{2}}$ at the UCEP gives $\tilde{h}=\delta\tilde{B}_{y}+\frac{1}{2}\tilde{\alpha}^{2}$, $c_{2}=2\delta\tilde{B}_{y}$ and $c_{1}=8(2-\tilde{g}^{2})/\tilde{g}^{2}$. The susceptibility is given by $\chi=\frac{\partial^{2}f}{\partial \tilde{h}^{2}}$. With $\tilde{h}\sim \tilde{\alpha}^{2}$, it leads to the logarithmic singularity $\chi\sim2\ln{\tilde{\alpha}}^{2}$ at $\tilde{\alpha}=0$ when $c_{2}=0$.  In Fig.~\ref{fig-fig3}(a), when the free energy of the local minimum at $\tilde{\alpha}\neq 0$ is identical to $f(\tilde{\alpha}=0)$ (the blue-green line), for $c_{2}\rightarrow 0^{-}$ (the yellow line), $\tilde{\alpha}=0$, the system is approaching the logarithmic divergence (marked by the black dot), indicating a continuous phase transition. While for $c_{2}\rightarrow 0^{+}$ (the purple line), $\tilde{\alpha}$ locates at the local minimum, acting as a first-order transition, without divergence. 

When $c_{1}$ increases from the UCEP, the local minimum lifts from $f=0$. Then, when we change $c_{2}$, the system first undergoes a continuous phase transition (induced by the nonanalytical point), followed by a first-order phase transition (induced by the local minimum). In contrast, when $c_{1}$ decreases, the local minimum covers the nonanalytical point, and only the first-order phase transition happens. If $c_{1}$ continuously increases, the local minimum shrinks, leading to the TCP. The above discussion gives the phase diagram of Eq.~\eqref{eq-UCEP}, depicted in Fig.~\ref{fig-fig3}(b), where the phase diagram has three phases with one first- and two continuous phase transitions with a UCEP and a TCP, matching the phase diagram in Fig.~\ref{fig-fig1} (a). The analysis gives the UCEP at $c_{1}=1/e$, $c_{2}=0$ and the TCP at $c_{1}=2/e^{\frac{3}{2}}$, $c_{2}=1/e^{\frac{3}{2}}$. With $c_{2}=2\delta\tilde{B}_{y}$ and $c_{1}=8(2-\tilde{g}^{2})/\tilde{g}^{2}$, it leads to $(\tilde{B}_{y},\tilde{g})=(1,4\sqrt{\frac{e}{1+8e}})\approx(1,1.38278)$ for the UCEP, and $(\tilde{B}_{y},\tilde{g})=(1+\frac{1}{2e^{3/2}},2e^{3/4}\sqrt{\frac{2}{1+4e^{3/2}}})\approx(1.11157,1.37635)$ for the TCP. Compared to the results in Fig.~\ref{fig-fig1} (a), point B ($1, 1.38288$), point C ($1.10022, 1.37698$), the predictions of Eq.~\eqref{eq-UCEP} are accurate, with tiny deviations resulting from the assumption $\tilde{\alpha}\ll 1$ during the expansion in Eq.~\eqref{eq-point_B}. Technical details can be found in SM \cite{suppRao}.

The UCEP and unilateral criticality establish a qualitatively new class of phase transitions, fundamentally different from previous theories, such as multicriticality \cite{Chang1973Generalized,Hankey1972Geometric,Griffiths1975Phase,Tuthill1975Renormalization,Riedel1972Scaling,CHANG1992Differential,Griffiths1970Thermodynamics,Soriente2018Dissipation,Casta2012Universal,Overbeck2017Multicritical}, conventional critical endpoints \cite{Fisher1990Universality,Szydlowska2023Ferroelectric,poole1992phase}, and mixed-order phase transitions, in which order parameters act like a first-order transition while the correlation length diverges \cite{Bar2014Mixed,Thouless1969Long,Aizenman1988Discontinuity}. Many other models that may have a similar logarithmically singular form $f\sim\tilde{h}^{2}\ln{\vert \tilde{h}\vert}$ \cite{Wegner1972Corrections,Wegner1973Logarithmic}, such as the two-dimensional classical Ising model \cite{Wegner2016Two}, long-range Ising models at the Sak boundary \cite{Defenu2023Long,Koziol2021Quantum}, relativistic $1{+}1$-dimensional Dirac theories and the Gross-Neveu model \cite{Gross1974Dynamical}, and anisotropic Lifshitz $O(N)$ models at the upper critical dimension $d_{\mathrm{eff}}=d+z=4$ \cite{Diehl2001Liftshitz,diehl2002critical,Mergulh1998Field}. With coupling to a non-local operator \cite{ritsch2013cold,Mivehvar2021Cavity}, $\tilde{h}\sim c_{2}+\tilde{\alpha}^{2}$, the system will match the form of Eq.~\eqref{eq-UCEP} with model details only changing $c_{1}$ and $c_{2}$. Hence, the UCEP and TCP could happen. Thus, the unilateral criticality is not a model-specific peculiarity, representing a new, potentially universal type of phase transitions, likely realized in a broad class of systems.

(6) Within the mean-field theory, in terms of the two order parameters, the ground-state energy density reads
$
    \epsilon_{\text{MF}} = \tilde{\gamma}\tilde{\alpha}^{2}+m_{s}^{2}-\sqrt{\tilde{\alpha}^{2}+4m_{s}^{2}+\tilde{B}_{y}^{2}} \,. 
$
It still has three phases (PSR, AFN, PN), but the UCEP disappears and transition lines intersect at point C; see SM \cite{suppRao}.

\begin{figure}[htbp]
	\centering
 \includegraphics[width=0.48\textwidth]{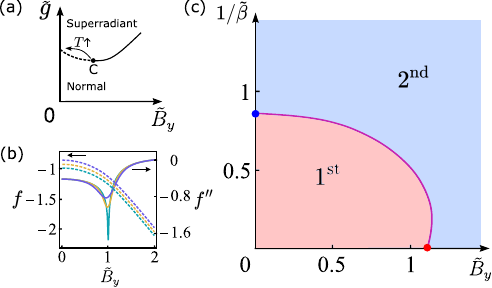}
\caption{(a) The schematic phase diagram at finite $T$, with only a TCP (point C). As $T$ increases, point C will move to the vertical axis and eventually vanish for $T>T_c$. (b) The free energy density $f=F/(NJ)$ (dashed) and its second derivative $f''$ (solid) for $\alpha=0$ when $\tilde{\beta} = 5$ (purple), 10 (yellow), and 100 (blue-green). (c) The phase diagram showing whether the normal-superradiant phase transition is of first- or second-order in the $\tilde{\beta}$-$\tilde{B}_y$ plane, where $\tilde{\beta} \equiv J/(k_BT)$. The blue dot on the vertical axis is at $\tilde{\beta}_c = 1.14299$ (or $k_B T_c = J/1.14299$). The red dot corresponds to point C in Fig.~\ref{fig-fig1}(a).}
\label{fig-fig4}
\end{figure}

{\it Finite-temperature effect}: The free energy at $T$ is 
\begin{equation}
F = \omega\alpha^{2}-\frac{N}{2\pi}\int_{-\pi}^{\pi}dk {\frac{1}{\beta}} \ln 2\cosh({\beta}\Omega_k),
    \label{eq-Eg1finiteT}
\end{equation}
where $ \Omega_{k}$ is in Eq.~\eqref{eq-Eg1}. For $T>0$, the zero-temperature quantum phase transition between AFN and PN will become a crossover \cite{Sachdev2011Book}. Hence, the finite-temperature phase diagram contains only two phases, as shown in Fig.~\ref{fig-fig3}(a) and (b), and the UCEP does not exist at finite temperature. Fig.~\ref{fig-fig3}(c) shows, for given $T$ and $B_y$, whether the normal-superradiant phase transition is first- or second-order when $g$ varies across the critical value. There exists a threshold temperature $T_c$, indicated by the blue dot on the vertical axis. For $T>T_c$, the transition is always second-order. In contrast, for $T<T_c$, both first- and second-order transitions exist. As a result, the TCP survives at a finite temperature for $T<T_c$. See detailed discussion in SM \cite{suppRao}.

{\it Discussion and Conclusion}: We propose two physical systems to realize the above models. The first is based on superconducting qubits that interact with a transmission line resonator \cite{majer2007coupling,luo2016dynamic,zhang2014quantum}, in which the interaction between neighboring sites can be realized by capacitive or inductive couplings. The second realization is based on Rydberg atoms in a cavity, which also has the above spin-spin interaction \cite{Dou2022Extended,Zhang2023Quantum,bacciconi2025local,Han2024Interaction,labuhn2016tunable,zeiher2016many} due to the long-range Coulomb interaction between neighboring atoms. As for the transverse field $B_y$, one may use a pair of phase-locked lasers that drive a two-photon transition from the electronic ground state $\vert g\rangle$ to a high-lying Rydberg state $\vert e\rangle$. The lasers are tuned near two-photon resonance and applied globally to all atoms. It leads to an effective transverse field for the two-level system \cite{labuhn2016tunable}.

To conclude, we report a unilateral critical endpoint in the zero-temperature phase diagram of the cavity-Ising model with $\mathbb{Z}_2$ symmetry in the thermodynamic limit. We construct the minimal description of the UCEP, which also gives a TCP. It indicates that the unilateral criticality is not model-specific and can be realized in a wide range of systems. Experimental realization appears feasible based on existing platforms. In recent years, the coupling between cavity and superfluids \cite{Zheng2020CavityFFLO}, atomic cloud\cite{sauerwein2023engineering}, Bose-Hubbard model \cite{Dogra2016Phasetransitions}, superconductivity \cite{Curtis2019CavitySC, Schlawin2019CavitymediatedSC}, quantum Hall effect \cite{Felice2022Breakdown, Enkner2025Tunable}, and solid materials \cite{Schlawin2022CavityMaterial,wang2024spin,eisenach2021cavity} have been intensively explored.  We expect that our discovery of unilateral criticality and the relation between symmetry and phase transitions can provide general guidelines for extending these phenomena to new regimes and platforms. The realization of this new criticality may have applications in fields such as quantum metrology \cite{SunZhe2010FisherInformation, Cheng2025SuperHeisenberg, Rams2018AttheLimits, Alushi2024Optimality} and quantum sensing \cite{Degen2017Quantum}. 

\textit{Acknowledgments}: This work is supported by the Strategic Priority Research Program of the Chinese Academy of Sciences (Grant No. XDB0500000), National Natural Science Foundation of China (No. U23A2074, and No. 12275203), and the Innovation Program for Quantum Science and Technology (2021ZD0301200, 2021ZD0301500). HP acknowledges support from U.S. NSF and the
Welch Foundation (Grant No. C-1669). X.-W. Luo is also supported by USTC start-up funding. 

\bibliography{ref.bib}

\begin{thebibliography}{121}
\bibitem [{Sig()}]{SignofJ}%
  \BibitemOpen
  \href@noop {} {}\bibinfo {note} {When $J < 0$, the antiferromagnetic-normal (AFN) phase will turn to a ferromagnetic-normal (FN) phase, and the whole phase diagram is unchanged. The free energy $F$ is independent of the sign of $J$.}\BibitemShut {Stop}%
\end{thebibliography}

\end{document}